## 1.1 Machine-Detector Interface

Nikolai V. Mokhov, Fermilab, Batavia, IL, USA
Mail to: mokhov@fnal.gov

### 1.1.1 Introduction

In order to realize the high physics potential of a Muon Collider (MC) a high luminosity of $\mu^+\mu^-$-collisions at the Interaction Point (IP) in the TeV range must be achieved (~$10^{34}$ cm$^{-2}$s$^{-1}$). To reach this goal, a number of demanding requirements on the collider optics and the IR hardware, arising from the short muon lifetime and from relatively large values of the transverse emittance and momentum spread in muon beams that can realistically be obtained with ionization cooling [1], should be satisfied. These requirements are aggravated by limitations on the quadrupole gradients [2] as well as by the necessity to protect superconducting magnets and collider detectors from muon decay products [3, 4]. The overall detector performance in this domain is strongly dependent on the background particle rates in various sub-detectors. The deleterious effects of the background and radiation environment produced by the beam in the ring are very important issues in the Interaction Region (IR), detector and Machine-Detector Interface (MDI) designs. This Chapter is essentially based on studies presented very recently [5, 6].

### 1.1.2 Sources of Background and Radiation Load

There are three sources of beam-induced backgrounds and radiation loads in MC: incoherent $e^+e^-$ pair production at the interaction point, beam halo loss on limiting apertures and muon beam decays. The first source, with its 10-mb cross-section, gives rise to backgrounds in detector of $3\times10^4$ electron pairs per bunch crossing. It can be confined with the appropriate design of the MDI components assisted by the high solenoidal field of the detector. The second source is taken care of by the beam halo collimation (extraction) far upstream of IR.

Muon decays is the major source. At 0.75-TeV, with $2\times10^{12}$ muons in a bunch, one has $4.28\times10^5$ decays per meter of the lattice in a single pass, and $1.28\times10^{10}$ decays per meter per second for two beams. Electrons from muon decays have mean energy of approximately 1/3 of that of the muons. These ~250-GeV electrons, generated at the above rate, travel to the inside of the ring magnets, and radiate a lot of energetic synchrotron photons tangent to the electron trajectory. Electromagnetic showers induced by these electrons and photons in the collider components generate intense fluxes of muons, hadrons and daughter electrons and photons, which create high background and radiation levels both in a detector and in the storage ring at the rate of 0.5-1 kW/m (to be compared to a few W/m in hadron colliders). Without corresponding mitigation measures, the quench stability and cryogenic issues in superconducting magnets and background loads to a detector have a potential of killing the idea of a high-energy muon collider.

### 1.1.3 IR Lattice

The basic parameters of the muon beams and of the collider lattice necessary to achieve the desired luminosity are given in a previous Chapter. The major problem to solve was correction of the chromaticity of IR quadrupoles in such a way that the dynamic aperture remained sufficiently large and did not suffer much from strong beam-beam effect. To achieve these goals a new approach to the IR chromaticity correction was developed [1] which may be called a three-sextupoles scheme. The IR layout and beam sizes for $\beta^* = 1$ cm are shown in Fig. 1.

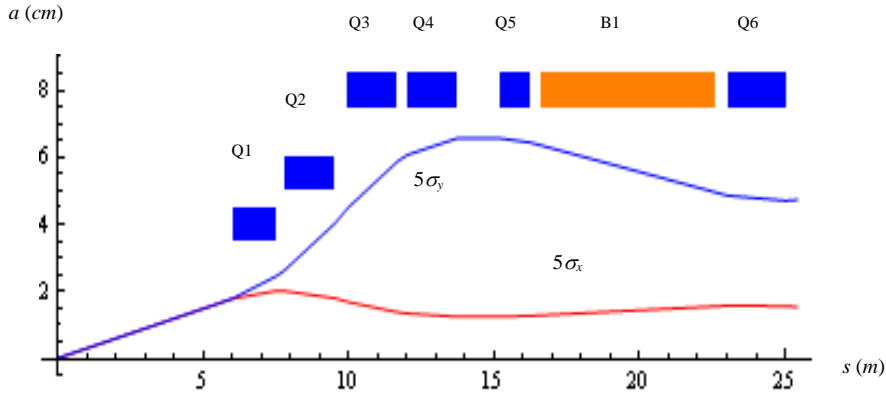

**Figure 1:** Beam sizes and aperture of the final focus.

### 1.1.4 IR Magnet Design

The IR doublets are made of relatively short quadrupoles (no more than 2 m long) to optimize their aperture according to the beam size variation and allow placing protecting tungsten masks in between them. The first two quadrupoles in Fig. 1 are focusing ones and the next three are defocusing ones. The space between the 4$^{th}$ and 5$^{th}$ quadrupoles is reserved for beam diagnostics and correctors. The cross-sections of MC IR quadrupoles feature two-layer shell-type $Nb_3Sn$ coils and cold iron yokes. The coil aperture ranges from 80 mm (Q1) to 160 mm (Q3 to Q5). The nominal field in the magnet coils is ~11-12 T, whereas the maximum field reaches ~13-15 T. As can be seen, all the magnets have ~12% margin at 4.5 K, which is sufficient for the stable operation with an average heat deposition in magnet mid-planes up to 1.7 mW/g. Operation at 1.9 K would increase the magnet margin to ~22% and their quench margin by a factor of 4.

The specifics of the heat deposition distributions in the MC dipoles – with decay products inducing showers predominantly in the orbit plane – require either a very large aperture with massive high-Z absorbers to protect the coils or an open midplane design [1-3]. It has been shown [7] that the most promising approach is the open mid-plane design which allows the decay electrons to pass between the superconducting coils and be absorbed in high-Z rods cooled at liquid nitrogen temperatures, placed far from the coils. The coils are arranged in a cos-theta configuration [1, 7]. The coil aperture in the IR dipoles is 160 mm, the gap height is 55 mm with supporting Al-spacers, and magnetic length is 6 m. The nominal field is 8 T.

### 1.1.5 Energy Deposition in IR Magnets

Energy deposition and detector backgrounds are simulated with the MARS15 code [8]. All the related details of geometry, materials distributions and magnetic fields for lattice elements and tunnel in the ±200-m region from IP, detector components [9], experimental hall and machine-detector interface (Fig. 2) are implemented in the model. To protect the superconducting magnets and detector, 10 and 20-cm tungsten masks with 5 $\sigma_{x,y}$ elliptic openings are placed in the IR magnet interconnect regions and a sophisticated tungsten cone inside the detector [3, 4] were implemented into the model and carefully optimized. The 0.75-TeV muon beam is assumed to be aborted after 1000 turns. The cut-off energy for all particles but neutrons is 200 keV, while neutrons are followed down thermal energies (~0.001 eV).

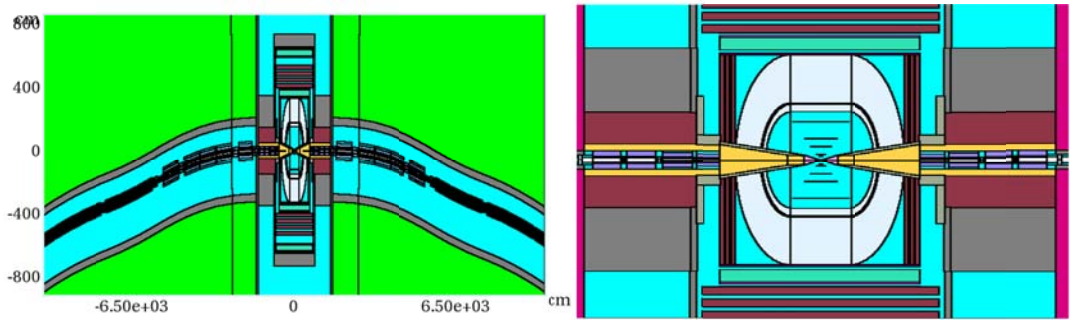

**Figure 2:** MARS15 model of IR (left) and MDI (right).

Calculated power density (absorbed dose) profiles are shown in Fig. 3 for the second and fifth final focus quadrupoles and in Fig. 4 for the first IR dipole. The right side in these plots is toward the ring center, the peak energy deposition is on this side for the IR dipoles and defocusing quadrupoles.

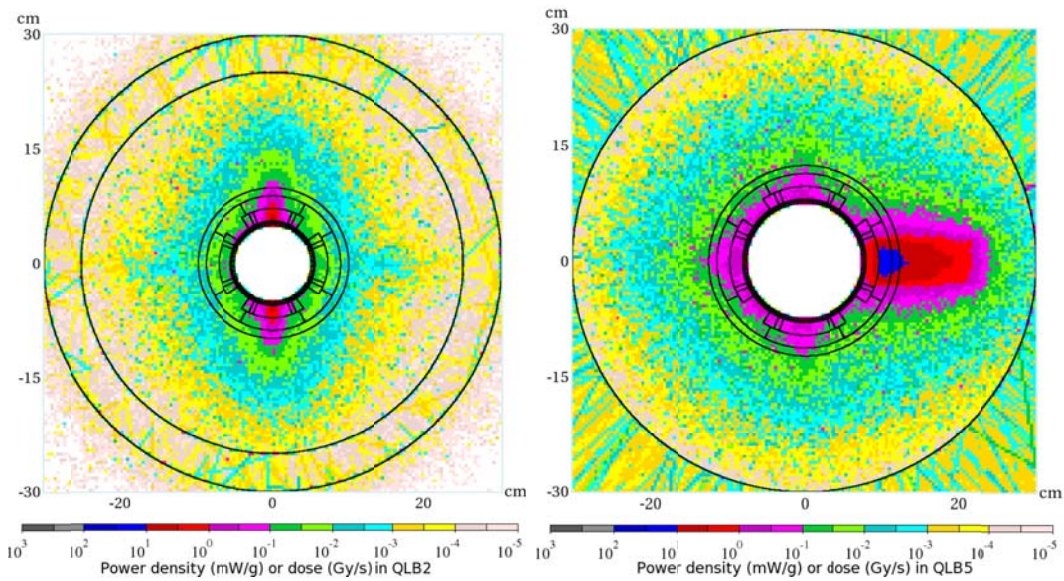

**Figure 3:** Power density (absorbed dose) profiles in the QLB2 focusing quadrupole (left) and QLB5 defocusing quadrupole (right).

The open midplane design for the dipoles provides for their safe operation. The peak power density in the IR dipoles is about 2.5 mW/g, being safely below the quench limit for the $Nb_3Sn$ superconductor-based coils at the 1.9-K operation temperature. At this temperature, first four quadrupoles are operationally stable, while the level in the next three IR quadrupoles is 5 to 10 times above the limit. This heat load could be reduced by a tungsten liner in the magnet aperture.

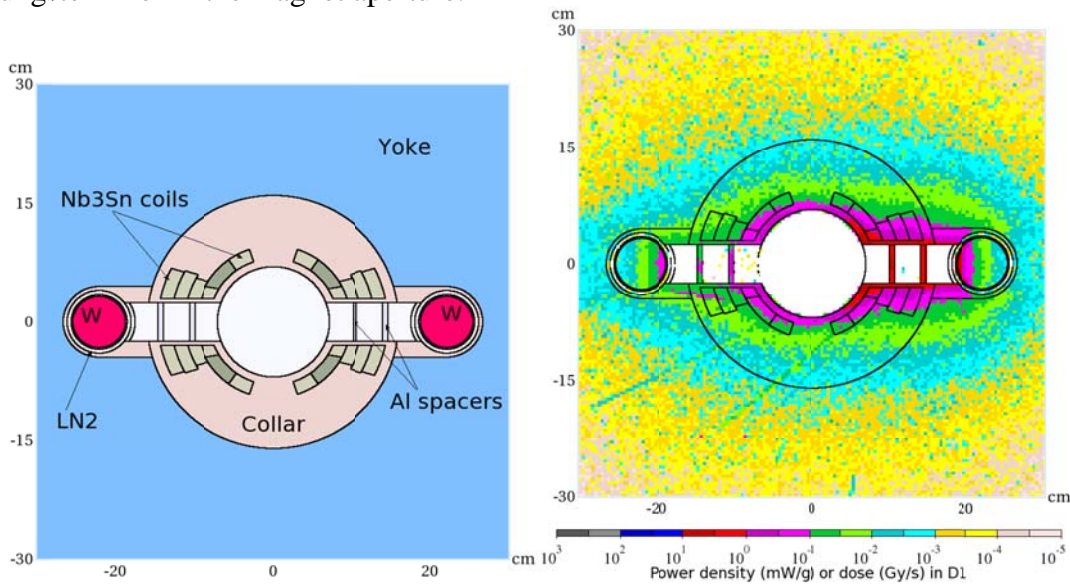

**Figure 4:** MARS15 model of the first IR dipole (left) and power density profiles in it (right).

### 1.1.6 MDI and Detector Backrounds

In the IR design assumed, the dipoles close to the IP and tungsten masks in each interconnect region (needed to protect magnets) help reduce background particle fluxes in the detector by a substantial factor. The tungsten nozzles with 10-20 degree outer angle in the 6 to 600 cm region from the IP (as proposed in the very early days of MC [10] and optimized later [1,3]), assisted by the detector solenoid field, trap most of the decay electrons created close to the IP as well as most of incoherent $e^+e^-$ pairs generated in the IP. With sophisticated tungsten, iron, concrete and borated polyethylene shielding in the MDI region (Fig. 2), total reduction of background loads by more than three orders of magnitude is obtained. With such an IR design, the major source of backgrounds in detector is muon decays in the region of about ±30 m from the IP.

Fig. 5 (left) shows muon flux isocontours in the MC IR. Note that the cut-off energy in the tunnel concrete walls and soil outside is position-dependent and can be as high as a few GeV at 50-100 m from the IP compared to 0.2 MeV close to the IP. These muons – with energies of tens to hundreds of GeV - illuminate the entire detector. They are produced by energetic photons from electromagnetic showers generated by decay electrons in the lattice components. The neutron isofluences inside the detector are shown in Fig. 5 (right). The maximum neutron fluence and absorbed doses in the innermost layer of the silicon tracker for a one-year operation are at a 10% level of that in the LHC detectors at the nominal luminosity. More work is needed to further suppress the very high fluences of photons and electrons in the tracker and calorimeter which exceed those at proton colliders.

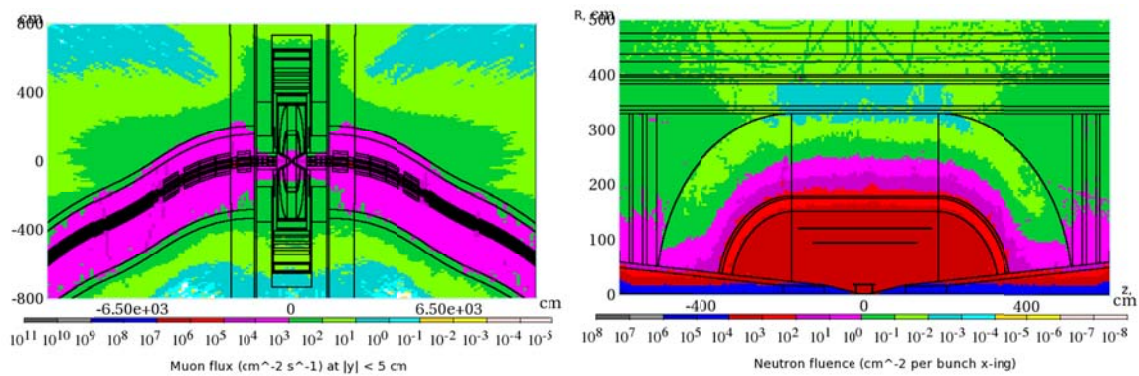

**Figure 5:** Muon isofluxes in IR (left) and neutron isofluences in the detector (right).

Realistic simulations of detector performance in presence of these backgrounds have started and revealed that backround loads are manageable thanks to all the measures implemented into the MDI design. It was found that with fast (< 0.5 ns) vertex and tracker Si detectors and time gate of 2-3 ns, one can get further substantial reduction of number of background-induced hits.

Acknowledgements to my co-authors Y.I. Alexahin, V.V. Kashikhin, S.I. Striganov, N.K. Terentiev and A.V. Zlobin.